\documentclass[preprint,showpacs,preprintnumbers, amsmath,amssymb]{revtex4} 

\usepackage{graphicx}\graphicspath{./graph/} 
\usepackage{bm}
\usepackage{color}
\usepackage{verbatim}
\usepackage{amssymb}
\usepackage{amsfonts}
\usepackage{latexsym}
\usepackage{epstopdf}

\definecolor{red}{rgb}{1,0,0}
\definecolor{green}{rgb}{0,1,0}
\definecolor{blue}{rgb}{0,0,1}

\newcommand{\be}{\begin{equation}}
\newcommand{\ee}{\end{equation}}
\newcommand{\bea}{\begin{eqnarray}}
\newcommand{\eea}{\end{eqnarray}}
\newcommand{\bdm}{\begin{displaymath}}
\newcommand{\edm}{\end{displaymath}}

\newcommand\ptl{\partial}

\begin{document}

\title{Concerning Energy}

\author{T. Erber}
\affiliation{Department of Physics Illinois Institute
of Technology, Chicago\\
Department of Applied Mathematics Illinois Institute
of Technology, Chicago, USA\\
Department of Physics University of Chicago, Chicago
, USA \\
\date{submitted 3 Aug 2019}  \\
{\rm Dedicated to the memory of Ilya Prigogine}}

{

\vspace{20mm}

\begin{abstract}
Energy has an ambiguous status in general relativity. For  systems embedded in asymptotically flat space-times it is possible to construct an integral invariant that corresponds to total energy, however there is no local differential invariant that can be identified with energy density. Moreover, in cosmological 'big-bang' scenarios there is an energy gain of about 70 orders of magnitude between the initial detonation and final inflation. Nevertheless, there is a widespread belief that all physical systems, irrespective of their size or complexity, can be associated with a unique scalar measure --- their energy. 
This presumption suggests parallels with the mathematical concept of measure in set theory as well as with entropy.
However, both analogies are limited in scope. We also discuss a wide variety of other forms of energy ranging from particle physics to information theory, and consider the implications for conservation laws. Finally, we recall several historical episodes in which energy conservation was at the 
center of controversy.
\end{abstract}

\maketitle

\section{Energy and general relativity}

The assertion that energy is not conserved in general relativity is tucked away in an article on ``Particle Physics and the Cosmic Microwave Background'' that appeared in a recent issue of a mainline physics journal 'Physics Today' ([1],  p.30). Similar remarks are scattered throughout the general relativity literature but in a more muted form and partially obscured by the complicated formalism of tensor calculus. The energy ambiguities are inherent in the very basis of general relativity, namely the principle of equivalence. It is presumed that in every infinitesimal world region in which the space and time variations of gravity can be neglected there always exists a coordinate system in which  gravitation has no influence either on the motion of particles or any other physical process. In other words, in any infinitesimal world region every gravitational field can be transformed away ([2], p.145). This is a formal restatement of the original surmise concerning the apparent absence of gravitational effects
in a freely falling elevator. In recent times this thought experiment has been implemented on a large scale in a NASA program to accustom astronauts to weightless conditions by flying them in an aircraft following a parabolic trajectory in the earth's atmosphere. According to the equivalence principle this simulation is physically indistinguishable from the imaginary situation in which the astronauts are removed to a remote region of space far away from any masses so that the ambient gravitational field is negligibly small ([3], p.15). The unavoidable conclusion is that if there is no gravitational field, then there is no corresponding local gravitational energy density. Nevertheless it will appear later that it is possible to construct a global integral invariant that can be identified with the total energy.

In  general relativity the effects of gravity are incorporated in the curvature of space-time. In order to understand the energy problem at a slightly more technical level it is necessary to introduce a minimal amount of differential geometry and a slight touch of tensor calculus. Let $u_1, u_2, u_3, u_4$ represent the curvilinear coordinates of a point in a 4-dimensional space-time reference system $K$, and $\vec{r}$ a 4-component vector extending from the origin of the $K$ system to that point. The standard notation for the dependence of $\vec{r}$ on the $u$-coordinates is
\be
\vec{r}=\vec{r}(u_1,u_2,u_3,u_4)\ . \label{rofu}
\ee
This can be confusing because the symbol $\vec{r}$ on the lhs simply denotes the vector whereas the $\vec{r}$ on the rhs actually represents a set of 4 functions that map the coordinates $u_1,\cdots$ onto the 4 components of the vector $\vec{r}$. In order to obtain the curvature information specified by the $K$-system it is expedient  to compute the differential of arc length $ds$. As usual this is determined by the square $ds^2$ which is given by the 4-dimensional dot product
\be
ds^2=d\vec{r}\cdot d\vec{r}\   \label{ds2}
\ee
where from (\ref{rofu})
\be
d\vec{r}=\frac{\ptl\vec{r}}{\ptl u_1}du_1 +\frac{\ptl\vec{r}}{\ptl u_2}du_2 +\frac{\ptl\vec{r}}{\ptl u_3}du_3+\frac{\ptl\vec{r}}{\ptl u_4}du_4\ .  \label{dr}
\ee
Written out explicitly (\ref{ds2}) and (\ref{dr}) lead to a lengthy expression
\bea
	&\hspace{-10mm}& ds^2=\frac{\ptl\vec{r}}{\ptl u_1}\cdot\frac{\ptl\vec{r}}{\ptl u_1}du_1^2+\frac{\ptl\vec{r}}{\ptl u_1}\cdot\frac{\ptl\vec{r}}{\ptl u_2}du_1du_2+\frac{\ptl\vec{r}}{\ptl u_1}\cdot\frac{\ptl\vec{r}}{\ptl u_3}du_1du_3+
\frac{\ptl\vec{r}}{\ptl u_1}\cdot\frac{\ptl\vec{r}}{\ptl u_4}du_1du_4\nonumber\\
&&\hskip .9cm+\frac{\ptl\vec{r}}{\ptl u_2}\cdot\frac{\ptl\vec{r}}{\ptl u_1}du_2du_1+\frac{\ptl\vec{r}}{\ptl u_2}\cdot\frac{\ptl\vec{r}}{\ptl u_2}du_2^2+\frac{\ptl\vec{r}}{\ptl u_2}\cdot\frac{\ptl\vec{r}}{\ptl u_3}du_2du_3+
\frac{\ptl\vec{r}}{\ptl u_2}\cdot\frac{\ptl\vec{r}}{\ptl u_4}du_2du_4\nonumber\\
&&\hskip .9cm+\frac{\ptl\vec{r}}{\ptl u_3}\cdot\frac{\ptl\vec{r}}{\ptl u_1}du_3du_1+\frac{\ptl\vec{r}}{\ptl u_3}\cdot\frac{\ptl\vec{r}}{\ptl u_2}du_3du_2+\frac{\ptl\vec{r}}{\ptl u_3}\cdot\frac{\ptl\vec{r}}{\ptl u_3}du_3^2+
\frac{\ptl\vec{r}}{\ptl u_3}\cdot\frac{\ptl\vec{r}}{\ptl u_4}du_3du_4\nonumber\\
&&\hskip .9cm+\frac{\ptl\vec{r}}{\ptl u_4}\cdot\frac{\ptl\vec{r}}{\ptl u_1}du_4du_1+\frac{\ptl\vec{r}}{\ptl u_4}\cdot\frac{\ptl\vec{r}}{\ptl u_2}du_4du_2+\frac{\ptl\vec{r}}{\ptl u_4}\cdot\frac{\ptl\vec{r}}{\ptl u_3}du_4du_3+
\frac{\ptl\vec{r}}{\ptl u_4}\cdot\frac{\ptl\vec{r}}{\ptl u_4}du_4^2\ \;  .\nonumber\\\label{ds2long}
\eea
Encountering this unwieldy formula at the outset of the calculations makes it clear that some notational simplifications are needed. Accordingly we first introduce an abbreviation for the dot product of the vector derivatives appearing in (\ref{ds2long}), viz.
\be
 \frac{\ptl\vec{r}}{\ptl u_i}\cdot\frac{\ptl\vec{r}}{\ptl u_j}=g_{ij}\ ,\quad i,j=1,2,3,4\ . \label{gij}
\ee
Then, using the standard notation for sums, (\ref{ds2long}) can be rewritten in a condensed form
\be
ds^2=\sum_{i=1}^4\sum_{j=1}^4g_{ij}du_idu_j\ . \label{ds2compact}
\ee
Further, since these kinds of expressions occur over and over again, it is convenient to eliminate the summation signs by using the Einstein summation convention. This simply says that two repeated indices in an expression are automatically to be summed over. Using this convention (\ref{ds2long}) can finally be written in the concise form
\be
ds^2=g_{ij}du_idu_j\ . \label{ds2einstein}
\ee
All the information concerning the curvature of space-time is now incorporated in the $g_{ij}$ functions which form the basis for most of the subsequent development of general relativity. It can be shown that the $g_{ij}$'s are components of a covariant tensor of rank 2; because of its central role it is often called the fundamental tensor or the metric tensor. The transformation properties of tensors guarantees that the squared distance (\ref{ds2einstein}) remains invariant under a change in coordinate systems. Furthermore, there are only 10 independent coefficients in the quadratic form (\ref{ds2einstein}) because of the index symmetry $g_{ij}=g_{ji}$ implied by (\ref{gij}).

In the complete absence of gravity (\ref{ds2einstein}) reduces to the 'flat' space-time of special relativity
\be
ds^2=dx_1^2+dx_2^2+dx_3^2-dx_4^2\ . \label{flat}
\ee
The first three terms correspond to the spatial components of an orthogonal cartesian coordinate system, and the fourth is given by $x_4=ct$ following Minkowski's convention for the space-time of special relativity. Comparing (\ref{ds2einstein}) and (\ref{flat}) it is clear that in this case the components of the metric tensor are given by
\be
g_{ij}=1 \quad{\rm for}\quad i=j=1,2,3\ ; \quad g_{44}=-1\ ;\quad g_{ij}=0\quad {\rm for}\quad i\ne j\ . \label{flatout}
\ee

Space-time can be slightly curved by introducing a weak quasi-static gravitational field. To consider the coupling to a particle at rest, we need consider  only the $g_{44}$ component of the metric tensor.  One  obtains
\be
g_{44}=-1-\frac{2\phi}{c^2}\ ,\label{g44mod}
\ee
where $\phi$ is the ordinary scalar potential of Newtonian gravity and $c$ is the velocity of light ([2], p.151). The energy density of the gravitational field in Newtonian gravity is given by
\be
\frac{1}{8\pi}\big|\nabla\phi\big|^2 \ .   \label{newtenergy}
\ee
This form is plausible because the gradient of the potential is the gravitational field, and, in analogy with electromagnetism, the square of the field is  proportional to the energy density. Since (\ref{g44mod}) shows that $\phi$
is linked to $g_{44}$, the most likely candidate for the energy density of the gravitational field in general relativity is an expression involving the squares of the first derivatives of the metric tensor. This supposition leads to the crux of the energy density problem. As emphasized by Wald:  ``$\cdots$ since no tensor other than $g_{ij}$ itself can be constructed locally from only the coordinate basis components of $g_{ij}$ and its first derivatives, a meaningful expression quadratic in the first derivatives of the metric can be obtained only if one has additional structure in space-time''. However, since general realativity presumes that the metric components $g_{ij}$ fully describe all aspects of space-time no additional structure is necessary ([4], p.286). This antinomy implies that there is no local expression for the gravitational energy density analogous to the Newtonian formula.

Despite the lack of a local expression for gravitational energy density, a global integral invariant for the total energy can be constructed. A simplified derivation of this result starts from the ordinary divergence theorem
\bea
&&\int \nabla\cdot\vec{A}~dV=\int\vec{A}\cdot\vec{n}~dS\ , \label{div}\\
&&\hskip .52cm{\rm volume}\hskip 1.4cm{\rm surface}\nonumber
\eea
which relates the properties of a vector field $\vec{A}$ throughout the volume $V$ with its behavior on a boundary surface $S$. If we were to loosely identify the integrand on the lhs with the gravitational energy density --- which we just argued doesn't exist --- then it seems that nothing would be gained by shifting from volume to surface integrals because the rhs of (\ref{div}) wouldn't exist either. Einstein found a way to restore useful meaning to this situation by assuming that the material bodies were arranged in an 'island' configuration: that is a set of isolated bodies where space-time is curved only in finite regions whereas at remote locations it is asymptotically nearly flat ([5]). In this case the boundary surface in (\ref{div}) can be expanded to lie only in the nearly flat region of space-time, and the corresponding tensor version of the rhs of (\ref{div}) can be reduced to the form
\be
\frac{1}{16\pi}  \sum_{i,j = 1}^{3}  \int_S\bigg
(\frac{\ptl g_{ij}}{\ptl x_i}  - 
\frac{\ptl g_{ii}}{\ptl x_j}  \bigg) dS_{j} \ .\label{reduced}
\ee
The contact with physics can now be established by specializing to the situation where there is only a single central body with mass $M$, with the convention $G= 1$ and $c = 1$ . The asymptotically almost 'flat' value of the space-space components of the  metric tensor then are
\be
g_{ij}\approx \delta_{ij} \bigg( 1+2\frac{M }{ r} \bigg) \ ,
\ee
where  $r$  is the radial component of a spherical coordinate system centered at $M$ and $\delta_{ij}= 1, i = j $ and vanishes for $i \neq j$. With these simplifications (\ref{reduced}) can be evaluated to yield $Mc^2$, which confirms its identification with the total energy ([3], p.48).

\section{Energy and Entropy}

It is generally taken for granted that energy and entropy are part of the immutable foundations of physics with almost unrestricted ranges of application. But in practical situations the determination of the entropy $S$ is limited by severe restrictions. The entropy associated with the transition between two states can be defined only if there exists at least one reversible process that connects these states. Indeed such reactions are the staples of books on physical chemistry. However, completely different conditions prevail in processes such as the plastic deformation of materials, the cumulation of microscopic damage in fatigue, or the acceleration of structural failure in stress-corrosion. These are situations where, in Bridgman's terms, the states are completely surrounded by irreversibility so that it is impossible to leave the initial state by any path whatsoever that is not irreversible in  detail ([6], p.56).
Consequently, the classical entropy concept is applicable only in a highly idealized set of conditions and is irrelevant in some of the commonest situations of daily life.

Other difficulties arise from the fact that entropy is a collective measure that cannot be derived from an underlying Hamiltonian mechanics. It was first suggested by Poincar\'e that in the context of Hamiltonian mechanics it is impossible to construct a function $S$ of the generalized coordinates and momenta that increases monotonically along orbits ([7], 1889). This was proved rigorously with modern methods by Olsen ([8], 1993). Since the entropy accompanying physical processes is always non-decreasing these results show that thermodynamics cannot be based on purely mechanical principles. An alternative approach is to introduce combinatorics and probabilistic assumptions ([9], 1959). The central result is an expression that relates entropy to the logarithm of the probability $W$ of the occurrence of a state, viz.
\be
S=k\ln W\,\label{entln}
\ee
where $k$ is Boltzmann's constant. The persistent problems associated with entropy now assume a different form. For instance, a supercooled liquid is not in a state of internal thermodynamic equilibrium: its state is not determined solely by phenomenological variables such as temperature and pressure. In fact such systems can exist in many possible microscopic configurations, and according to (\ref{entln}) all with different  entropies. These systems cannot be treated thermodynamically because thermodynamics is concerned principally with the entropy of the most probable state. In Simon's view one simply cannot apply thermodynamics to these types of systems in any way ([10], p.16).

The measurement of energy is generally more robust than the determination of entropy. Complete irreversibility is not an impediment: for instance,  the conversion of mechanical work into heat by friction was noted by Benjamin Thompson ([11], p.490) as early as 1800. The wide acceptance of the energy concept rests on numerous experiments demonstrating the interconversion of energy in many areas including physics, chemistry, and engineering. Already in 1879 Rowland listed eight different kinds of energy whose transformations into one another furnish the mechanical equivalent of heat: (a) mechanical energy; (b) heat; (c) electrical energy; (d) magnetic energy; (e) gravitational energy; (f) radiant energy: (g) chemical energy; (h) capillary energy ([12], p.405). He followed this up by a synopsis of dozens of experimental methods for obtaining the energy equivalents among the forms of energy just cited ([12], p.407).

 The developments occurring in subsequent years are described in many references including the books by Hiebert [13], Elkana [14], and Coopersmith [15] and Jaffe and Taylor [31]. A recent issue of the American Journal of Physics was entirely devoted to energy [32]. Among the current views of energy conservation there is at least one tinged with a hint of sarcasm. It is argued that since energy has no precise definition any proposed violation of its conservation could always be swept aside by a suitable redefinition. By such drastic means the conservation of energy is reduced to a sterile tautology ([16], p.30). Fortunately, recent discoveries of new forms of energy have led to a more fruitful perspective. We briefly consider these cases:

\subsection{Energy and particle physics}

For many purposes a nucleus can be viewed as a collection of nucleons (neutrons and protons) held together by a `nuclear force', and its rest energy (equivalently its mass) is the total rest energy of the nucleons minus a binding energy. In modern particle theory the nucleons are themselves composites of quarks and gluons. The analog of the binding energy is now a much more subtle concept, since the quarks and gluons are `confined', never appearing as directly observable free particles.

For over a century numerous sub-atomic particles have been identified each with its own characteristic rest energy. The most recent is the Higgs boson found at the CERN LHC accelerator in 2012. Although this object is referred to as an elementary particle its rest energy of 120 GeV makes it roughly $2.4\times 10^5$ more massive than the electron.

\subsection{Energy and information}

Computer installations are usually equipped with cooling systems designed to dissipate heat generated by currents flowing through thousands of circuit elements. In 1991 Landauer suggested that the information stored and processed by physical systems would itself also have an energy equivalent. By considering a simple information storage  device --- a system with two distinct states coupled to a reservoir at temperature $T$ --- Landauer showed that the work associated with the creation of a single bit, i.e., a binary choice, was of the order $kT\ln 2$, where $k$ is Boltzmann's constant. A similar estimate applies to the erasure of a single bit of information. Although this `quantum' of information is extremely small --- $kT\ln 2\sim 3\times 10^{-21}$ Joule at room temperature --- it does constitute a new form of energy ([17],  p.30).

In cases where the quantification of energy in terms of bits is straightforward, there is no difficulty in extrapolating Landauer's results to more general situations. But there is uncertainty associated with cryptography where the basic aim is to disguise information by encipherment, diffusion, and pre-arranged coding procedures. The situation is even murkier in cases where information cannot be rendered in digital form. The status of the `new' energy then remains problematic.

\subsection{Dark energy}
The most intriguing candidate for a new kind of energy is dark energy. Speculations concerning this mysterious entity were prompted by the discovery that astronomical objects near the periphery of the visible universe had an outward acceleration. To date there is no tangible evidence supporting either the existence of dark energy or its conceptual partner dark matter. But elaborate experimental efforts are underway at CERN and other laboratories to detect weakly interacting massive particles (WIMPs) and very weakly interacting sub-EV particles (WISPs)  which may or may not indicate the presence of dark constituents in the universe. These unfettered speculations have created an opening for unconventional ideas concerning violations of energy conservation as a source of dark energy. Two controversial options are non-unitary modifications of Schr\"odinger's wave equation and quantum gravity theories that invoke a fundamental discreteness in the structure of space-time [18].

\section{Energy and measure theory}

Given any physical system, regardless of its size or complexity --- it may be a photon, a proton, or a cluster of galaxies --- it is presumed that it can be associated with a unique scalar measure, its energy. In general, measures satisfy several criteria: (i) If a set $A$ is cut up into a finite number of disjoint sets that are reassembled to form a set $B$, then $A$ and $B$ have the same measure; (ii) The measure of the union of disjoint sets is the sum of their individual measures. Specifically, if '$m$' denotes a measure, and $\bigcup$ and $\bigcap$ are the union and intersection operations, then
\be
m(A\bigcup B)=m(A)+m(B)\quad {\rm where}\quad  A\bigcap B=\emptyset\ ;
\ee
and by extension
\be
m(\bigcup_i A_i)=\sum_i m(A_i)\quad{\rm when}\quad A_i\bigcap A_j =\emptyset\quad{\rm for}\quad i\ne j\ .
\ee
These criteria are broad enough to cover both the mathematical notions of measure corresponding to length etc., as well as the physical concept of energy. In the mathematical physics literature these distinctions are usually glossed over although  obviously there is no resemblance between a physical system with internal interactions and a mathematical set that may contain an uncountable number of elements. A conspicuous distinction arises from the existence of non-measureable sets. In particular, the Banach-Tarski theorem implies that 3-dimensional or `solid' sets of this type can be decomposed and reassembled to to form another solid set of any specified shape and volume ([19], p.2; [20], Chapter 5). It is an open question whether there are any analogous anomalies associated with energy. In an Appendix to his book, Wagner mentions a speculative resemblance between the Tarski-Banach construction and the `big bang' expansion of the universe. In a practical sense the issue is moot because physical systems of arbitrary complexity have an indeterminate energy content.

\section{Energy gone astray}

There are three prominent episodes in the history of physics where energy conservation was at the center of controversy.

\subsection{Lord Kelvin and the age of the earth}

During the course of the nineteenth century religious chronologies that dated the earth's beginnings to about 5300 BCE came into open conflict with the much longer time scales implied by Darwin's theory of evolution. Evidence from geology also pointed in the direction of terrestrial time scales extending into many millions of years. Lord Kelvin (William Thompson) entered the fray near the beginning of his scientific career in 1846 and maintained a lifelong interest in estimating the age of the earth. Unlike many of his peers he had a command of higher mathematics and was one of the founders of modern thermodynamics. In 1890 he became President of the Royal Society. Ironically, it was the combination of his scientific proficiency and unquestioned prestige as a leading British physicist that led to his unbending defense of his estimates regarding the age of the earth and blinded him to new possibilities.

In brief, Kelvin shifted the focus of lifetime estimates from the earth to the sun. Knowing the geometry of the earth-sun system and the amount of light incident on the earth --- circa $0.6 \, \mbox{cal/(cm$^2$ sec)}$ --- it is possible to deduce the  loss of radiant energy by the sun: about $3.86 \,\times \, 10^{26}$ Watt. Kelvin estimated that if the entire mass of the sun were consumed by the most energetic chemical process known, the energy loss could not be sustained for more than 3000 years. In casting about for alternate sources of energy, Kelvin chanced on the idea of meteoritic impact. A crude estimate of the energy available from this process can be obtained by assuming that an aggregate of meteors with a total mass equal to the sun's mass is initially dispersed throughout a very large spherical shell centered on the sun. If the shell implodes onto the sun owing to gravitational attraction, the total energy released on impact is sufficient to supply the heat for 20 million years.

Of course this estimate still clashed with the  much longer time scale required by biological and geological arguments, but Kelvin held fast to this number up to the time of his death in 1907. The logical possibility that these disputes could be resolved by the introduction of a new form of energy apparently did not occur to Kelvin [21]. In fact, the connection between radioactivity and the solar age problem was established swiftly in 1903. Pierre Curie  and Albert Laborde provided quantitative data on the amount of heat emanating from radium [22], and within weeks William E. Wilson proposed in a letter to Nature that this new form of energy powered the sun ([23], [24]).

\subsection{Is energy conservation violated in quantum mechanics?}

Quantum mechanics had a long gestation period, beginning with Planck's black body spectrum in 1900 and extending to the introduction of matrix mechanics and wave mechanics in 1925-6. In the intervening years the tremendous amount of experimental data accumulated from the study of atomic and nuclear phenomena was organized with the help of numerous empirical formulas derived from semi-classical quantization rules. Most of those fell into disuse after the advent of the 'real' quantum mechanics in 1926; but one important result that survived was Einstein's determination of the ratio of induced to spontaneous emission in atomic radiation processes [25].

Our concern is with another paper with the ambitious title ``The Quantum Theory of Radiation'' published by Bohr, Kramers, and Slater in 1924 ([26]). They write: ``At the present state of science it seems necessary, as regards the occurrence of transition processes, to content ourselves with considerations of probability. Such considerations have been introduced by Einstein, who has shown how a remarkably simple deduction of Planck's law of temperature radiation can be obtained $\cdots$'' [25]. Unfortunately, these remarks misconstrue Einstein's reasoning. In actuality, he used consistency with the pre-existing Planck law to deduce the ratio of spontaneous to induced emission, and also reconfirmed the basic relation between energy changes in atomic level transitions and the frequency of the emitted photon,
\be
E({\rm initial})-E({\rm final})=\hbar\omega\ .
\ee
In developing this theory Einstein was always careful not to confound probabilistic reasoning with any possibility of violating energy conservation. But this was not the path followed by Bohr {\it et al.}. They claim ``...(it) would seem the only consistent way of describing the interaction between radiation and atoms (is) by a theory involving probability considerations. This independence reduces not only conservation of energy to a statistical law, but also conservation of momentum.'' [26]. Within two years this confusing detour was replaced by a comprehensive quantum theory of radiation based on wave mechanics. Strict energy conservation was incorporated into this formalism from the outset.

\subsection{Energy conservation leads to the discovery of neutrinos}

$\beta$-decay is the process in which an electron or positron is emitted by a nucleus. Experiments show that the emitted particles have a broad single-peaked energy distribution with a maximum energy given by the difference in energy between the initial and final states of the nuclei undergoing $\beta$-decay. This observation confirms that energy isn't created, but still leaves open the question of what happened to the missing energy. In 1931 Pauli made the bold suggestion that conservation of energy could be restored if an additional hard-to-detect particle --- the forerunner of today's WISPs --- was emitted during the $\beta$-decay. According to this proposal, the transformation of a neutron into a proton would be an interaction with three objects in the final state,
\be
n \rightarrow p+e^-+\nu_e\ ,  \label{betadk}
\ee
where $\nu_e$ denotes a neutrino [27]. The variable sharing of  energy between the electron and the neutrino accounts for the broad electron energy spectrum. Charge conservation implied by eq.(\ref{betadk}) shows that the neutrino is a neutral object. Precise measurements of the maximum electron energies also indicate that the neutrino rest mass is very small. Finally, rough estimates of the neutrino interaction cross-sections with matter yield values of the order of $10^{-44} \; {\rm cm}^2$, or 20 orders of magnitude smaller than the usual nuclear cross sections. One striking consequence of this feeble interaction is that neutrinos should be able to pass through massive objects like the sun with little probability of being absorbed. Despite these daunting parameters, advances in experimental techniques finally enabled Reines and Cowan to claim the direct detection of free neutrinos in 1955 [28]. Since that time neutrino physics has progressed considerably; currently it is one of the most active areas in all of physics.

\subsection{A contemporary energy puzzle}

Consider an inelastic one-dimensional collision where a mass $m_0$ with initial velocity $v_0$ impinges on another mass $M$ initially at rest. Suppose the two masses coalesce and that the composite object $m_0+M$ moves away from the collision point with velocity $v_f$. Then an elementary calculation using momentum conservation shows that during the collision the total kinetic energy decreases by an amount given by
\be
\frac{1}{2} \bigg(\frac{m_0 \,M}{m_0 + M}\bigg)\ v_0^2. \label{dke}
\ee
The trite answer to the question 'where did the missing energy go?' is that it was converted to heat. Recently this dismissive attitude has been challenged by a number of studies of elastic collision schemes where heat plays no apparent role and yet energy seems to disappear in mysterious ways.

A canonical example has been discussed by Johnson and Atkinson [29]. They consider an infinite sequence of masses $(m_0,m_1,m_2,\cdots)$ aligned in order along a straight line. The initial mass $m_0$ is projected toward the first mass $m_1$ with a velocity $v_0$; it is presumed that $m_1$ and all the other masses are initially at rest. After $m_0$ collides elastically with $m_1$, $m_0$ acquires a velocity $V_0$ and $m_1$ moves forward towards $m_2$ with velocity $v_1$. This collisional round continues until all subsequent masses have undergone two collisions.  Assuming that all masses recoil with the same speed $V~(=V_0)$ simplifies the calculation but still preserves the essential energy relations. In this case it is convenient to introduce a dimensionless parameter $\lambda$ where
\be
\lambda = \frac{v_0}{V} >1\ .   \label{lambda}
\ee
Some further computations then show that the kinematics constrains the individual mass ratios as follows:
\be
\frac{m_n}{m_0}=\frac{\lambda(\lambda-1)}{(\lambda+n)(\lambda+n-1)}\ .
\ee
This sequence converges with sufficient rapidity to show that the total mass of the entire system is finite,
\be
\sum_{n=0}^\infty m_n=\lambda m_0\ .
\ee
After all the collisions have been completed the final kinetic energy of all the masses is
\be
\frac{m_0 \, v_0 \, V}{2} .
\ee
This value is less than the initial energy $m_0v_0^2/2$. A similar result is obtained if the kinematics are described by special relativity.  As emphasized by Atkinson and Johnson, it remains a mystery how an infinite sequence of purely elastic collisions could possibly mimic an inelastic process. Without heat, where is the energy?

\section{Final reflections}

Ernst Mach devoted an entire book to the conservation of energy [30]. All the essential arguments follow from the basic premise ``that one cannot create  something out of nothing''. This was believed without exception for a long time until the advent of the `big bang scenario' for the creation of the universe. According to the `grand concordance' model of astrophysics the entire universe resulted from a singular quantum fluctuation of the vacuum that was amplified
and dispersed at super-light velocities by an `inflaton' field [1]. When these revolutionary notions were first put forward by George Gamow about 70 years ago they were met by disbelief, derision, and contempt. Nowadays these unfamiliar ideas
are taken for granted, although it is probable that the paradigm will change as more astrophysical data accumulate. In the meanwhile, the different aspects of energy cited in the preceding sections make it extremely unlikely that we will ever arrive at a precise and general definition of energy. Further, the amorphous nature of energy casts doubt on the
feasibility of devising a general proof of its conservation. In the foreseeable future we will probably have to be content with a more modest program: following the pattern set by Rowland [12], in every novel situation (dark energy!) the conservation of energy will have to be verified anew on a case-by-case basis.

\section{Acknowledgement}

It is a pleasure to thank my colleague Professor Porter Johnson for constructive criticism and help with this manuscript.

\end{document}